# Applying Agile Requirements Engineering Approach for Re-engineering & Changes in existing Brownfield Adaptive Systems

Abdullah Bin Masood[1] and Muhammad Asim Ali[2]

[1] Student at Mohammad Ali Jinnah University

[2] Assistant Professor at SZABIST, Karachi Campus

**Abstract.** Requirements Engineering (RE) is a key activity in the development of software systems and is concerned with the identification of the goals of stakeholders and their elaboration into precise statements of desired services and behavior. The research describes an Agile Requirements Engineering approach for re-engineering & changes in existing Brownfield adaptive system. The approach has few modifications that can be used as a part of SCRUM development process for re-engineering & changes. The approach illustrates the re-engineering & changes requirements through introduction of GAP analysis & requirements structuring & prioritization by creating AS-IS & TO-BE models with 80 / 20 rule. An attempt to close the gap between requirements engineering & agile methods in form of this approach is provided for practical implementation.

## 1. Introduction

Most Organizations in this era have existing large or medium scale operational systems and with the evolvement of business some requirements also arise for change in existing system to meet evolving business demands & needs. In general, the challenges faced by industries are mainly the capture of requirements for changes & re-engineering in operational systems. Unfortunately, there are very limited approaches defined for re-engineering & changes in existing Brownfield operational systems because most of the system & software requirements literature assume development of system from scratch i.e. Green Field Systems.

A common scenario of Brownfield system development can be a major upgrade in current operational system in terms of following requirements

- Incorporation of new business rules in existing system.

- Adding up new feature in existing system.

- Up-gradation of existing feature.

- Adoption of new technology.

- Legalization / Certification of product or specific feature.

While fulfilling the above mentioned requirements most of the Brownfield projects do suffer from the same issues which are faced by Greenfield such as missed deadlines, cost overruns, compromised goals, too many change requests, feature & scope creeps but above all they do face unique challenges such as intertwined & associated modules, undocumented business constraints, obscure organizational memory & difficult to handle non-functional requirements in understanding of existing system. On top of that there are pressures from multiple stakeholders in terms of resistance to change justification of re-engineering in context of past investments and re-use of existing assets.

## 2. Challenges & Difficulties of Re-engineering & changes in Brownfield Systems

Unfortunately, there is no specialized model designed for requirements engineering which provides comprehensive strategy to handle above mentioned problems in re-engineering of Brownfield adaptive systems. Unlike Greenfield Requirements Engineering approach, while going through with the re-engineering and change in existing system, there are some obvious questions arise in one's mind which are categorized below

## 3. Analysis of existing & proposed system

1. What is the limitation of current system for which we are going through with the change?
2. Do we have enough knowledge of current system? (Undocumented system features)
3. To what extent should existing as well as new features be documented?

Structure & Prioritize existing & proposed requirements

1. What effects will this change have on other sub systems and modules? (Intertwined modules)
2. How to structure & prioritize requirements?
3. How interdependent requirements will be managed?

## 4. Verification & Validation of Functional & Non-functional requirements

1. What if re-engineered system misses out some critical undocumented business requirement which is ambiguous in current existing system?
2. How changes & re-engineering will be verified?
3. How non-functional requirements will be captured for re-engineered system?

## 5. Survey Questionnaire & Results

Data has been collected with a mixture of closed & open questionnaire / interviews from various industry professionals to validate the above mentioned problem. A consolidated result is shown in table 1.

**Table 1 Consolidated Survey Results**

| Re-Engineering & Change Impacts | Response on each impact category | | | |
|---|---|---|---|---|
| | Inbox Business Technologies | Avanza Soultions | Quality Check Inc | PrisLogix |
| **Requirements Model / Approach** | | | | |
| Existing Model / Approach used | No | No | No | No |
| Model / Approach followed by company | No | No | No | No |
| Need of Agile Model | Yes | Yes | Yes | Yes |
| **Re-engineering / Changes** | | | | |
| Incorporation of new feature. | Yes | Yes | Yes | No |
| Adding up new feature | No | Yes | Yes | No |
| Up-gradation of existing feature | Yes | Yes | Yes | Yes |
| Adoption of new technology | Yes | No | Yes | No |
| Legalization / Certification of product or specific feature. | No | No | Yes | No |
| **Problems & Challenges** | | | | |
| Requirements churn | No | Yes | Yes | No |
| Missed deadlines | Yes | No | Yes | No |
| Too many change requests | Yes | No | Yes | Yes |
| Scope / Feature creeps | Yes | Yes | Yes | No |
| Undocumented business constraints | Yes | Yes | Yes | Yes |
| Tightly coupled modules | No | Yes | Yes | No |
| Miss out critical feature | Yes | Yes | Yes | Yes |
| Difficult Non-Functional Requirements | No | No | No | Yes |
| **Impact of Re-engineering Problems** | | | | |
| Indefinite testing | No | Yes | Yes | No |
| Problem in other modules | Yes | Yes | Yes | Yes |
| **Business Needs** | | | | |
| Cross functional teams | Yes | Yes | Yes | Yes |
| Understanding of Business Domain | Yes | Yes | Yes | Yes |

**Requirements Engineering in Agile Methods**

In this section we will not provide the details of Requirements Engineering but in order to fill this gap we will discuss the detail of Agile Methodology applied on our problem. In contrast to traditional Requirements Engineering, Agile Methods do rely on 'Customer' for elicitation, elaboration & negotiation of requirements. Requirements are incepted, elicited & elaborated mostly through 'User Stories' & Class Responsibility Collaborator (CRC) cards with the help of co-located customer. Agile methods are characterized by Time Boxed Iterations which has prioritized requirements catered from User Stories. In effect, the early versions of the iterations become prototype & these prototypes are then refined in next iterations. In this way client becomes responsible for getting the requirements right. There are no incomplete, inconsistent and unambiguous requirements but there are some obvious requirements for later discovery.

Generally, the above mentioned Requirements Engineering process is followed in all Agile Methodologies. From various Agile Methods we have chooses to apply **Scrum** on our problems mentioned in section 2 because of its more responsive nature to changing environment, predictive iterations with analysis & prioritization of requirements.

## 6. Understanding Scrum

Scrum development method was proposed in 1995 by Ken Schwber. It is based on the idea that many of the processes during project cannot be predicted therefore it addresses the processes in flexible way and utilizes small teams to deliver increments of software using "sprints". These sprints may last between 14-30 days, during which all parties strive to achieve a specific goal. No new requirements can be introduced during Sprints. Each day begins with a kick-off SCURM-meeting to assure productivity, effectiveness & quality control. A Product Backlog is maintained from which Items (Requirements) are selected for Sprint, after which the tasks on selected items are performed. It enables the project team to prioritize work based on actual business value which then leads to regular returns on investment and improved communication channels to management and product owners.

## 7. Applying SCRUM on Challenges & Difficulties of Re-engineering & changes

This section describes an approach for applying SCRUM in Re-engineering & changes of Brownfield adaptive systems. It should be taken in account that our approach is based on default SCRUM process with few modifications.

## 8. Gap Analysis

In SCRUM methodology, bunch of tasks are directly fed into the product backlog and tasks are selected for each sprint from product backlog and moved to development sprint backlog. In Greenfield projects, requirements are exactly focused on what is required, and the goal is to deliver a complete system which is verifiable against requirements. Contrary to Greenfield projects, Brownfield (Re-engineering) projects are not completely focused towards what is required but the **focus is to redo the implemented requirements in new ways**. Following critical questions arises from the very beginning of re-engineering project

1. What is the limitation of current system for which we are going through with the change?

2. Do we have enough knowledge of current system? (Undocumented system features)

3. To what extent should existing as well as new features be documented?

To discover the context of re-engineering & changes an obvious starting point would be to identify the gap between the existing functionality of system & desired changes required to be implemented. First step in this regard would be to develop **AS-IS** model which can define what business & functional requirements are fulfilled by existing Brownfield system. The next step is to identify the success criteria through individual & group of stakeholders who would also play the critical roles in preparation of AS-IS model. There would also be several other classes of success-critical stakeholder which would help us to fully define the features of existing system. These include but not limited to Customers, Marketing persons, customer support representative & clerical staff. These entire stakeholders would not only define the current model but they would also define the limitation of existing system and it would lead as to prepare our **TO-BE** model. The TO-BE model would define the complete re-engineering required in existing system and it would be used as a Product Backlog for each sprint. **It forms an alternate way of generating Product Backlog and do not follow the usual method of SCRUM.**

It is necessary to form **cross functional** team consists of different stakeholders including software developers and testers during the preparation of AS-IS & TO BE documents which would help in finding the objectives, goals, rationales, priorities & scenarios of each requirement documented in AS-IS & TO BE documents.

## 9. Structuring & Prioritizing requirements

The definition of each business process & requirements is performed during the GAP analysis without going into too much detail. Further refinement is performed by cross functional team before the start of each sprint. Each business process & requirement is decomposed into a detail description of desired functionality to ensure that all requirements are consistent, feasible, understandable & verifiable. Structuring & decomposition of requirements will lead to better feature sizing & change management that would ultimately figure out the actual deliverables of the sprint. After accomplishment of requirements refinement, the next phase is to sort out & prioritize the requirements which are inevitable & interdependent. For this purpose we would use the well know 80/20 rule to figure out those certain requirements which are inevitable for business operations. It generally states that 20 percent of a population or sample consumes 80 % of the resources, if we mold this rule in context of software development it would not be false that 80% of the people use 20% of the feature OR we can focus those features through which 80% of the business processes are covered i.e. 80% of the process covered by 20% of the features or vice versa as shown in figure1.

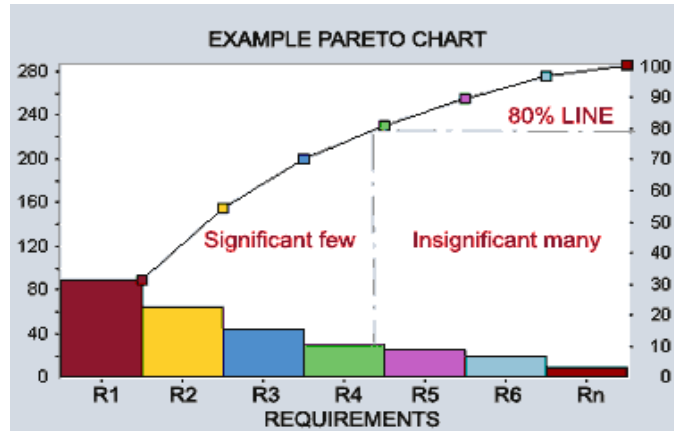
**Figure 1** Pareto chart for Requirements Prioritization

## 10. Re-engineering Sprint

Once requirements are structured, decomposed & prioritized, tasks are selected and physically moved into Sprint backlog, the 'real execution' section of the development cycle. There can be a single team or multiple teams which can participate in each sprint and distinct tasks are allocated to each team. At the start of sprint tasks are completely elaborated to all team members to ensure that each decomposed requirement is completely understood by all team members. Each working day starts with a scrum meeting during which the tasks performed by team members are discussed to improve the productivity & effectiveness of team. Problems faced by individuals regarding tasks are also discussed to identify the common issues faced by team which avoids the redundant work between different teams. No new requirements can be introduced during sprint to ensure the completion of sprint within estimated time. Any emerging change or new requirement found during sprint is inserted into product backlog and processed through same cycle of structuring & prioritizing but it is not necessary to go through with complete cycle for each requirement, if a requirement is already fine grained and obvious it can be directly prioritized and put into the next sprint backlog. A burn-down chart is maintained which described the amount of work accomplished to date by team. The end result of sprint is working deliverable which goes through the acceptance criteria to ensure that desired functionality has been achieved. The effectiveness of sprint is discussed at the end to suggest further improvements and to determine & prioritize existing and new requirements for next increment.

## 11. Re-engineering verification & validation

As mentioned in section 4.1 that it would be necessary to form cross functional team including testers for complete understanding of re-engineering & changes. The upfront scenario creation by testers will lead to improved feature coverage and verification of re-engineering impacts on other modules. Although the purpose of testing is to prevent defects but it also makes sure that all requirements mentioned are incorporated by creating requirements traceability matrix. Majority of questionnaire responses was in 'Yes' for omission of critical features during re-engineering. The early participation of testers will improve their domain knowledge and their creation of test cases in early stages will improve test coverage, change management & defect prevention activities. Testers will create test cases based on To

Be document created and aggregate them with their decomposed tasks. It would help testers to review initial ambiguities of requirements with domain experts & to validate requirements against sprint objectives with both aspects of functional and non-functional requirements. The overall strategy is to ensure that testing activities are integrated throughout the re-engineering cycle and to focus on the incorporation of requirements & their impacts on other modules. It would lead to defect prevention instead of defect detection which would be cost effective in all terms.

## 12. Validity Constraint & Future Work

The theory of research was established after discussion sessions with academic & market professionals. The goal of our research is to tailor an agile approach for requirement engineering of re-engineering & changes in existing Brownfield adaptive systems. Surveys were conducted in highly reputed organizations and their response was positive regarding need of approach. However, our study based on the theoretical aspects and no empirical study was performed for it. It would be very important while applying this approach practically on any re-engineering project to handle the operational aspects of systems because some features & operations of system are inevitable for business operations. We have just described how to include Re-engineering method in agile development process rather than describing the whole process model for Re-engineering & changes. The most important challenge would be to handle cultural, commercial & legal factors while analyzing the context of project. Secondly, It would also have to be taken in account that co-located customer does not understand the complete need of requirements and some requirements can be overlook by him/her. Finally this approach needs further refinement in terms of operational, social & commercial factors for implementation.

## 13. Conclusion

Recently there have been no approaches define to apply agile requirements engineering approach for re-engineering & changes in Brownfield adaptive systems. This study demonstrate a high level approach for re-engineering based on proven & well adopted agile method. The GAP analysis & requirements structuring activities in SCRUM approach can really be handy for re-engineering projects in terms of requirements documentation during AS-IS & TO BE formalization. It is really important to update the new processes & changes in TO BE document found during re-engineering sprint. The decomposition of requirements into tasks will help out in feature sizing & traceability links. Application of this approach to any practical project is necessary to find out the pros & cons of this approach.

## 14. References

[1] Boness, K. AND Harrison, R. 2007. Goal Sketching: Towards Agile Requirements Engineering. Software Engineering Advances 71-72

[2] Barry Boehm, 2009. Applying Incremental Commitment Model to Brownfield System Development. 7th Annual

Conference on System Engineering Research 2009 (CSER 2009)

[3] Srinivasan, J. AND Lundqvist, K. AND Malardalen Univ., Malardalen Using Agile Methods in Software Product Development: A Case Study. Information Technology: New Generations, 2009. ITNG '09. Sixth International Conference 1415 - 1420

[4] Tomayko, J.E. **Engineering of Unstable Requirements Using Agile Methods,** (2002)
[5] Vlaanderen, K. AND Brinkkemper, S. AND Jansen, S. AND Jaspers, E. The Agile Requirements Refinery: Applying SCRUM Principles to Software Product Management 1 – 10

[6] http://www.bcs.org/upload/pdf/complexity.pdf

[7] Requirements Engineering for Agile Methods. Alberto Sillitti, Giancarlo Succi – http://www.**agile**-itea.org/public/papers/sillitti-succiV2.pdf

[8] Agile Requirements engineering - http://www.infosys.com/infosys-labs/publications/setlabs-briefings/Pages/agile-methodology.aspx

[9] Book: Eating the IT Elephant: Moving from Greenfield Development to Brownfield [Richard Hopkins & Kevin Jenkins]

[10] James Chisan, Daniela Damian. Exploring the role of requirements engineering in improving risk management. In Proceedings of RE'2005. pp.481~482

[11] Daniela Damian, James Chisan. An Empirical Study of the Complex Relationships between Requirements Engineering Processes and Other Processes that Lead to Payoffs in Productivity, Quality, and Risk Management. IEEE Trans. Software Eng., 2006: 433~453

[12] Sabrina Marczak, Irwin Kwan, Daniela Damian Investigating Collaboration Driven by Requirements in Cross-Functional Software Teams Software Engineering Global interAction Lab – SEGAL University of Victoria, Victoria, Canada (smarczak, irwink, danielad)@cs.uvic.ca

[13] John Mylopoulos, Lawrence Chung, Brian A. Nixon. Representing and Using Nonfunctional Requirements: A Process-Oriented Approach. IEEE Trans. Software Eng., 1992: 483~497

[14] Douglas T. Ross, Kenneth E. Schoman Jr. Structured Analysis for Requirements Definition. IEEE Trans. Software Eng., 1977: 6~15

[15] Lawrence Chung, Julio Cesar Sampaio do Prado Leite. On Non-Functional Requirements in Software Engineering. In Proceedings of Conceptual Modeling: Foundations and Applications'2009. pp.363~379

[16] P. Berander, Evolving Prioritization for Software Product Management, ser. Blekinge Institute of Technology Doctoral Dissertation Series. Blekinge Institute of Technology, 2007.

[17] "Agile manifesto." [Online]. Available: http://www.agilemanifesto.org

[18] B. Fitzgerald, G. Hartnett, and K. Conboy, "Customising agile methods to software practices at intel shannon," European Journal of Information Systems, vol. 15, pp. 200–213, 2006.

[19] D. Greer and G. Ruhe, "Software release planning: an evolutionary and iterative approach," Information and Software Technology, vol. 46, no. 4, pp. 243–253, 2004

[20] Frauke Paetsch, Armin Eberlein, Frank Maurer. Requirements Engineering and Agile Software Development. In Proceedings of WETICE'2003. pp.308~313

[21] Shahzad Gul. A Comparative Study of Requirements Change Management in Non Agile and Agile Development Methodologies. In Proceedings of Software Engineering Research and Practice'2007. pp.115~121